# Creating and Probing Electron Whispering Gallery Modes in Graphene

Yue Zhao[1,2*], Jonathan Wyrick[1*], Fabian D. Natterer[1*], Joaquin Rodriguez Nieva[3*], Cyprian Lewandowski[4], Kenji Watanabe[5], Takashi Taniguchi[5], Leonid Levitov[3], Nikolai B. Zhitenev[1†], and Joseph A. Stroscio[1†]

[1]Center for Nanoscale Science and Technology, National Institute of Standards and Technology, Gaithersburg, MD 20899, USA
[2]Maryland NanoCenter, University of Maryland, College Park, MD 20742, USA
[3]Department of Physics, Massachusetts Institute of Technology, Cambridge, MA 02139, USA
[4]Department of Physics, Imperial College London, London SW7 2AZ, UK
[5]National Institute for Materials Science, Tsukuba, Ibaraki 305-0044, JAPAN
*These authors contributed equally to this work
†Corresponding authors. E-mail: nikolai.zhitenev@nist.gov, joseph.stroscio@nist.gov

**Abstract**: Designing high-finesse resonant cavities for electronic waves faces challenges due to short electron coherence lengths in solids. Previous approaches, *e.g.* the seminal nanometer-sized quantum corrals, depend on careful positioning of adatoms at clean surfaces. Here we demonstrate an entirely different approach, inspired by the peculiar acoustic phenomena in whispering galleries. Taking advantage of graphene's unique properties, namely gate-tunable light-like carriers, we create Whispering Gallery Mode (WGM) resonators defined by circular *pn*-junctions, induced by a scanning tunneling probe. We can tune the resonator size and the carrier concentration under the probe in a back-gated graphene device over a wide range, independently and *in situ*. The confined modes, revealed through characteristic resonances in the tunneling spectrum, originate from Klein scattering at *pn* junction boundaries. The WGM-type confinement and resonances are a new addition to the quantum electron-optics toolbox, paving the way to real-world electronic lenses and resonators.

Charge carriers in graphene exhibit light-like dispersion resembling that of electromagnetic waves. Similar to photons, electrons in graphene nanostructures propagate ballistically over micrometer distances, with the ballistic regime persisting up to room temperatures (*1*). This singles out graphene as an appealing platform for developing quantum electron-optics, which aims at controlling electron waves in a fully coherent fashion. In particular, gate-tunable heterojunctions in graphene can be exploited to manipulate electron refraction and transmission via the Klein scattering mechanism (*2*), in the same way as optical interfaces in mirrors and lenses are used to manipulate light. These properties have opened an avenue for exciting ideas in optics-inspired graphene electronics. First came Fabry-Pérot interferometers (*3*), which have been fabricated in planar *npn* heterostructures in single layer (*4*), and subsequently in bilayer (*5*) and trilayer graphene (*6*). The sharpness of the *pn* junctions achievable in graphene can enable precise focusing of electronic rays across the junction, allowing for electronic lensing and hyperlensing (*7–9*).

In this article, we report on electron Whispering Gallery Mode (WGM) resonators, a new addition to the quantum electron-optics toolbox. The WGM resonances are familiar for classical wave fields confined in an enclosed geometry−as happens, famously, in the whispering gallery of St. Paul's Cathedral. The WGM resonators for electromagnetic fields are widely used in a vast array of applications requiring high-finesse optical cavities (*10–12*). Optical WGM resonators do not depend on movable mirrors, and thus lend themselves well to high-Q designs. This, despite challenges in achieving tunability due to their monolithic (single-piece) character, often renders the WGM design advantageous over the Fabry-Pérot design [see *e.g.* (*12*) for a mechanically tunable optical WGM resonator]. Our system is unique from that perspective, being free from these limitations and representing a rare case of a fully tunable WGM resonator in which the cavity radius can be varied over a wide range by adjusting gate potentials. In contrast, the best electronic resonators known to date—the nanometer-sized quantum corrals designed by carefully positioning adatoms atop a clean metallic surface (*13*)—are not easily reconfigurable.

Further, while WGM resonators are ubiquitous in optics and acoustics (*10–12*, *14*), only a few realizations of such resonators were obtained in non-optical/acoustic systems. These include WGM for neutrons (*15*), as well as for electrons in organic molecules (*16*). Our scanning probe experiments are, to the best of our knowledge, the first direct demonstration of electron-based WGM resonators with a widely-tunable cavity. In our measurements we create a circular electron cavity beneath the tip of a scanning tunneling microscope (STM) and directly observe the WGM-type confinement of electronic modes. The cavity is defined by a tip-induced circular *pn* junction ring, at which the reflection and refraction of electron waves is governed by Klein scattering, see Fig.1. While Klein scattering results in perfect transmission and no reflection for normal incidence, it gives rise to nearly perfect reflection for oblique incidence occurring in the WGM regime (*2*). As illustrated in Fig.1B, this yields excellent confinement and high-finesse WGM resonances for modes with high angular momentum *m*, and a less perfect confinement for non-WGM modes with lower *m* values.

Electron optical effects in graphene have so far been explored using transport techniques, which lack spatial and angular resolution that would be indispensable for studying confined electronic states and/or electron lensing. The scanning probe technique, developed in this work, allows us to achieve nanometer-scale spatial resolution. In our experiment, the STM probe has dual purpose: creating a local *pn* junction ring which serves as a confining potential for electronic states, and probing by electron tunneling the resonance states localized in this potential. In our graphene device, the planar back gate and the STM tip, acting as a circular micrometer-sized top gate, can change both the overall background carrier density as well as the local carrier density under the tip. As such, *pn* and *np* circular junctions centered under the probe tip (Fig. 1A) can be tuned by means of the tip-sample bias, $V_b$, and back-gate voltage, $V_g$ [see Fig. S4 (*17*)]. Importantly, for the purpose of creating resonant electronic modes inside the junction, this configuration allows us to control *independently* and *in-situ* the carrier concentration beneath the STM tip as well as the *pn* ring radius. The tunneling spectral maps from such a device show a series of interference fringes as a function of the knobs $(V_b, V_g)$, see Fig. 2. As discussed below, these fringes originate from resonant quasi-bound states inside the *pn* ring.

The measured spacing between fringes can be used to infer the cavity radius. Using the formula $\Delta\varepsilon = \pi\hbar v_F/r$ and an estimate from Fig. 2A, $\Delta\varepsilon \approx 40$ meV, we obtain $r \approx 50$ nm, a value considerably smaller than the STM tip radius ($R \approx 1\mu$m). This peculiar behavior can be understood from a simple electrostatic model of a charged sphere proximal to the ground plane. For the sphere-to-plane distance *d* small compared to the sphere radius *R*, the induced image charge cloud behaves as $\rho(r) \propto 1/(d + r^2/2R)$, predicting a lengthscale $\sqrt{2Rd} \ll R$. This crude estimate is upheld, within an order of magnitude, by a more refined electrostatic modelling (*17*) which also gives a lengthscale much smaller than *R*.

The experimental results were obtained on a device consisting of a graphene layer on top of hexagonal boron nitride (*h*-BN), stacked on SiO$_2$ with a doped Si back gate [see supplementary materials for details (*17*)]. Figure 2A shows a tunneling conductance map as a function of back gate voltage on the horizontal axis ($V_g$) and sample bias ($V_b$) on the vertical axis. The striking feature seen in this map is a series of interference-like fringes forming a curved fan, labeled WGM', in the upper right of Fig. 2A. The center of the fan defines the charge neutrality point. This point can be off (0,0) in the $(V_g, V_b)$ plane due to impurity doping of graphene (shift along $V_g$) and contact potential difference between the probe tip and graphene (shift in $V_b$). As demonstrated in Fig. S5 (*17*), we shifted the center point of the fan to lower $V_b$ values by changing the tip work function with $D_2$ adsorption (*18*). Another interesting feature in such conductance maps is a (somewhat less visible) second fan of fringes, labeled WGM", which is crossing the primary WGM' fan. The fringes in the WGM" fan follow the typical graphene dispersion with respect to the Fermi energy, which varies with doping as $\propto \sqrt{V_g}$ from higher sample bias to lower as a function of $V_g$. Below we examine the primary (WGM') and secondary (WGM") fringes in more detail, confirming that they originate from the same family of WGM resonances.

Figure 2C shows 9 oscillations in a line cut across the WGM' fan along the $V_g$ axis. To understand the origin of these oscillations we examine the two spectral line cuts along the $V_b$ axis in Fig. 2D. The first spectrum in Fig. 2D at $V_g = -11$ V (blue curve) contains a group of resonances labeled 1" to 3", near the Fermi level ($V_b = 0$) with a spacing of (37.6±1.2) mV (*19*). In the map in Fig. 2A, these resonances can be seen to move to lower energies approximately following the typical Dirac point dispersion $\propto \sqrt{|V_g|}$. Taking a vertical cut at higher back gate voltage of $V_g = 16$ V (red curve) shows resonances 1" and 2" shifted down in energy in Fig. 2D. Focusing now at slightly higher energies, the WGM' resonances appear at positive energies in Fig. 2D, and are labeled 1' to 4' for $V_g = 16$ V. These four resonances are fit to Gaussian functions and shown deconvolved from the background conductance in the bottom right of the figure. The average spacing of these resonances is (116.9±7.5) mV (*19*). A close examination of Fig. 2A indicates the one-to-one correspondence between the WGM" resonances 1", 2",... and the WGM' resonances 1', 2'..., suggesting their common origin. Indeed, the WGM" resonances correspond to tunneling into the *pn* junction modes at energy $\varepsilon = \mu_0 + eV_b$, whereas the WGM' resonances reflect the action of the STM tip as a top gate, allowing tunneling into the same resonance mode at $\varepsilon = \mu_0$ [see Fig. S3 (*17*)]. For example, resonance 1" seen at $V_b \approx -100$mV is now accessible at the Fermi level by the tip-graphene potential difference, as shown in Fig. S3D (*17*), when tunneling into the WGM' resonance 1' at $V_b = 82$mV in Fig. 2A.

To clarify the WGM character of these resonances, we analyze graphene's Dirac carriers in the presence of a potential induced by the STM tip described by the Hamiltonian $H = H_0 + U(\mathbf{r})$, where $H_0$ is the kinetic energy term and $U(\mathbf{r})$ describes the STM tip potential seen by charge carriers. Since relevant length scales—the electron's Fermi wavelength and the *pn* ring radius—are much greater than the atomic spacing, we focus on the low-energy states. Using the *kp* approximation, we linearize the graphene electron spectrum near the *K* and *K'* points, bringing $H_0$ to the massless Dirac form: $\varepsilon \psi(\mathbf{r}) = (v_F \boldsymbol{\sigma} \cdot \mathbf{p} + U(\mathbf{r}))\psi(\mathbf{r})$, where $\mathbf{p} = -i\hbar \nabla_r$, and $\boldsymbol{\sigma} = (\sigma_x, \sigma_y)$ are pseudospin Pauli matrices. We take the tip potential to be radially symmetric reflecting the STM tip geometry. Furthermore, the distance from the tip to graphene, $d$, is considerably smaller than the electron's Fermi wavelength and the *pn* ring radius, both of which are smaller than the STM tip radius. We can therefore use a parabola to approximate the tip potential, $U(\mathbf{r}) = \kappa r^2$. The curvature $\kappa$, which affects the energy spectrum of WGM resonances, can be tuned with the bias and gate potentials, as discussed in the supplementary material (*17*).

The WGM states with different angular momentum can be described by the polar decomposition ansatz

$$\psi_m(r,\phi) = \frac{1}{\sqrt{r}} \begin{pmatrix} u_A(r)e^{i(m-1)\phi} \\ u_B(r)e^{im\phi} \end{pmatrix}, \quad (1)$$

with *m* an integer angular momentum quantum number, and *A, B* label sublattices. We non-dimensionalize the Schrödinger equation using the characteristic length and energy scales, $r_* =$ $\sqrt{Rd}$, $\varepsilon_* = \hbar v_F/\sqrt{Rd}$, to obtain the radial eigenvalue equation of the two-component spinor $u(r)$ with components $u_A(r)$ and $u_B(r)$:

$$\varepsilon\, u(r) = \left(-i\sigma_x \partial_r + \frac{m+1/2}{r}\sigma_y + \kappa r^2\right) u(r). \quad (2)$$

Here $r$ is in units of $r_*$, $\varepsilon$ is in units of $\varepsilon_*$, and $\kappa$ is in units of $\kappa_* = \varepsilon_*/r_*^2$. This equation is solved using a finite difference method, see supplementary material (*17*). We can use this microscopic framework to predict the measured spectral features. The tunneling current, expressed through the local density of states (DOS), is modeled as

$$I = \int_{\mu_0}^{\mu_0+eV_b} d\varepsilon\, T(\varepsilon, V_b) \sum_m D_m(\varepsilon), \quad D_m(\varepsilon) = \sum_{r,\nu} e^{-\frac{\lambda r^2}{2}} |u_\nu(r)|^2 \delta(\varepsilon - \varepsilon_\nu), \quad (3)$$

which is valid for modest $V_b$ values (*21*). Here $\mu_0$ is the Fermi energy under the tip which in general is different from ambient Fermi energy $\mu_\infty$ as a result of gating by the tip (see below). The quantity $T(\varepsilon, V_b)$ which depends on the tip geometry, work function and density of states, will be taken as energy independent. The quantity $D(\varepsilon) = \sum_m D_m(\varepsilon)$ represents the sum of partial-$m$ contributions to the total density of states beneath the tip, with $\nu$ labelling eigenstates of Eq. (2) with fixed $m$. The weighting factor $e^{-\lambda r^2/2}$ is introduced to account for the finite size of the region where tunneling occurs, with the Gaussian halfwidth $\lambda^{-1/2} \sim r_*$ [see supplementary material (*17*)].

The WGM resonances for different partial-$m$ contributions $D_m(\varepsilon)$, which combine into the total density of states, are shown in Fig. 3. Individual WGM states exhibit very different behavior depending on the $m$ value [see Fig. 3B and Fig.1B]. Klein scattering at the *pn* junction produces stronger confinement for the large-$m$ modes and weaker confinement for small-$m$ modes. Indeed, the Klein reflection probability is strongly dependent on the angle of incidence $\theta$ at the *pn* interface, growing as $R(\theta) \sim 1 - \exp[-\xi \sin^2(\theta)]$, where $\xi$ is a characteristic dimensionless parameter (*20*). The value of $\theta$ grows with $m$ as $\tan\theta \propto m$. As a result, larger values of $m$ must translate into larger reflectivity and stronger confinement. This trend is indeed seen clearly in Fig. 1B and Fig. 3B. Also, as *m* increases, mode wavefunctions are being pushed away from the origin, becoming more localized near the *pn* ring, in full accord with the WGM physics.

In order to understand how one family of WGM resonances gives rise to two distinct fans of interference features seen in the data, we must account carefully for the gating effect of STM tip. We start with recalling that the conventional STM spectroscopy probes features at energies $\varepsilon_\nu = \mu_0 + eV_b$, where $\varepsilon_\nu$ are system energy levels. This corresponds to the family WGM" in our measurements. However, as discussed above, the tip bias variation causes the Fermi level beneath the tip to move through system energy levels $\varepsilon_\nu$, producing an additional family of interference features (WGM') described by $\varepsilon_\nu = \mu_0$ [see Fig. S3 and accompanying discussion (*17*)]. To model this effect, we evaluate the differential conductance $G = dI/dV_b$ from Eq. (3) taking into account the dependence $\mu_0$ *vs.* $V_b$. This gives

$$G \propto (1-\eta)D(\mu_0 + eV_b) + \eta\, D(\mu_0), \quad (4)$$

with $\eta = -\partial\mu_0/\partial(eV_b)$. The two contributions in Eq. (4) describe the WGM' and WGM" families. We note that the second family originates from the small electron compressibility in graphene resulting in a finite $\eta$ and would not show up in a system with a vanishingly small $\eta$ (*e.g.* in a metal). We use Eq. (4) with a value $\eta = 1/2$ to generate Fig. 2B. In dyouoing so we relate the parameters ($\varepsilon$,$\kappa$) in the Hamiltonian, Eq. (2) and the experimental knobs $(V_b, V_g)$, using electrostatic modeling described in (*17*). This procedure leads to a very good agreement with the measure $dI/dV_b$, as illustrated in Figs. 2A and 2B.

In addition to explaining how the two sets of fringes, WGM' and WGM", originate from the same family of WGM resonances, our model accounts for other key features in the data. In particular, it explains the large difference in the WGM' and WGM" periodicities noted above. It also correctly predicts the regions where fringes occur, see Fig. 2B. The bipolar regime in which *pn* junction rings and resonances in the DOS occur, takes place for the probed energies ε of the same sign as the potential curvature. In the case of a parabola $U(r) = \kappa r^2$ this gives the condition $\varepsilon \kappa > 0$, corresponding to the upper-right and lower-left quadrants in Fig.3A inset. In experiment, however, the potential is bounded by $U(|r| \to \infty) = \mu_0 - \mu_\infty$ (see Fig.1A), which constraints the regions in which WGMs are observed (*17*). Accounting for the finite value $U(|r| \to \infty)$ yields the condition $|\varepsilon| \leq |\mu_0 - \mu_\infty|$, with $\text{sign}(\varepsilon) = \text{sign}(\kappa) = \text{sign}(\mu_0 - \mu_\infty)$. This gives the WGM' and WGM" regions in Fig.2B bounded by whited dashed and white dotted lines, respectively, and matching accurately the WGM' and WGM" location in our measurements.

The range of *m* values which our measurement can probe depends on the specifics of the tunneling region at the STM tip. We believe that a wide range of *m* values can be accessed, however, somewhat unfortunately, currently we are unable to distinguish different partial-*m* contributions, since the corresponding resonances are well aligned, see Fig.3. We note in this regard that different angular momentum *m* values translate into different orbital magnetic moment values, opening an interesting possibility to probe states with different *m* by applying a magnetic field.

Looking ahead, the above explanation of the observed resonances, in terms of the Whispering Gallery effect in circular *pn* rings acting as tunable electronic WGM resonators, has other interesting ramifications. First, it can shed light on puzzling observations of resonances in previous STM measurements (*22–24*), which hitherto remained unaddressed. Second, our highly tunable setup in which the electron wavelength and cavity radius are controlled independently lends itself well to directly probing other fundamental electron-optical phenomena such as negative refractive index for electron waves, Veselago lensing (*7*) and Klein tunneling (*2*). Further, we envision probing more exotic phenomena such as the development of caustics where an incident plane wave is focused at a cusp (*25–27*) and special bound states for integrable class of dynamics where the electron path never approaches the confining boundary at perpendicular incidence (*28*). These advances will be enabled by the unique characteristics of graphene which allow for electronic states to be manipulated at the microscale with unprecedented precision and tunability, thus opening a wide vista of graphene-based quantum electron-optics.

34. **Acknowledgements -** Y. Z. acknowledges support under the Cooperative Research Agreement between the University of Maryland and the National Institute of Standards and Technology Center for Nanoscale Science and Technology, Grant No. 70NANB10H193, through the University of Maryland. J.W. acknowledges support from the Nation Research Council Fellowship. F. D. N. greatly appreciates support from the Swiss National Science Foundation under project numbers 148891 and 158468. We thank Steve Blankenship and Alan Band for their contributions to this project, and we thank Mark Stiles and Phillip First for valuable discussions.

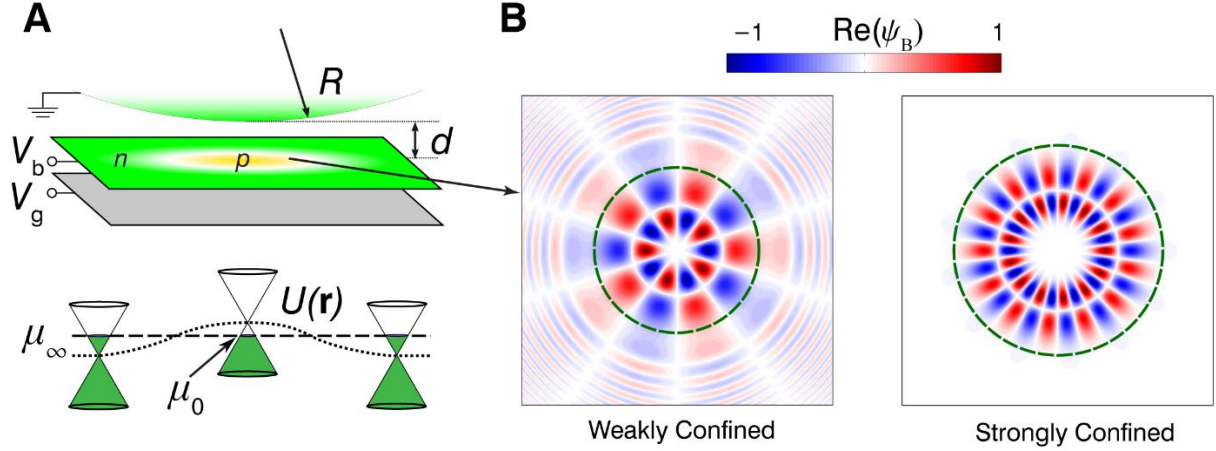

**Fig. 1. Confined electronic states in microscopic electron cavities defined by *pn* junction rings in graphene**. (**A**) The rings are induced by the STM tip voltage bias ($V_b$) and back-gate voltage ($V_g$) adjusted so as to reverse the carrier polarity beneath the tip relative to the ambient polarity. The *pn* junctions act as sharp boundaries giving rise to Klein scattering of electronic waves, producing mode confinement via the Whispering Gallery mechanism. The cavity radius and the local carrier density are independently tunable by the voltages $V_b$ and $V_g$. Electron resonances are mapped out by the STM spectroscopy measurements, see Fig. 2. Shown are the graphene band offset, the alignment induced by the STM tip potential ($r$), and the quantities discussed in the text: the STM tip radius ($R$), its distance from graphene ($d$), the local ($\mu_0$) and ambient ($\mu_\infty$) Fermi levels. (**B**) Spatial profile of WGM resonances. Confinement results from interference of the incident and reflected waves at the *pn* rings (dashed line). The confinement is stronger for the larger angular momentum *m* values, corresponding to more oblique wave incidence angles. This is illustrated for *m*=5 (weak confinement) and *m*=13 (strong confinement). Plotted is the quantity, Re($\psi_B$), the real part of the second spinor component in Eq. (1).

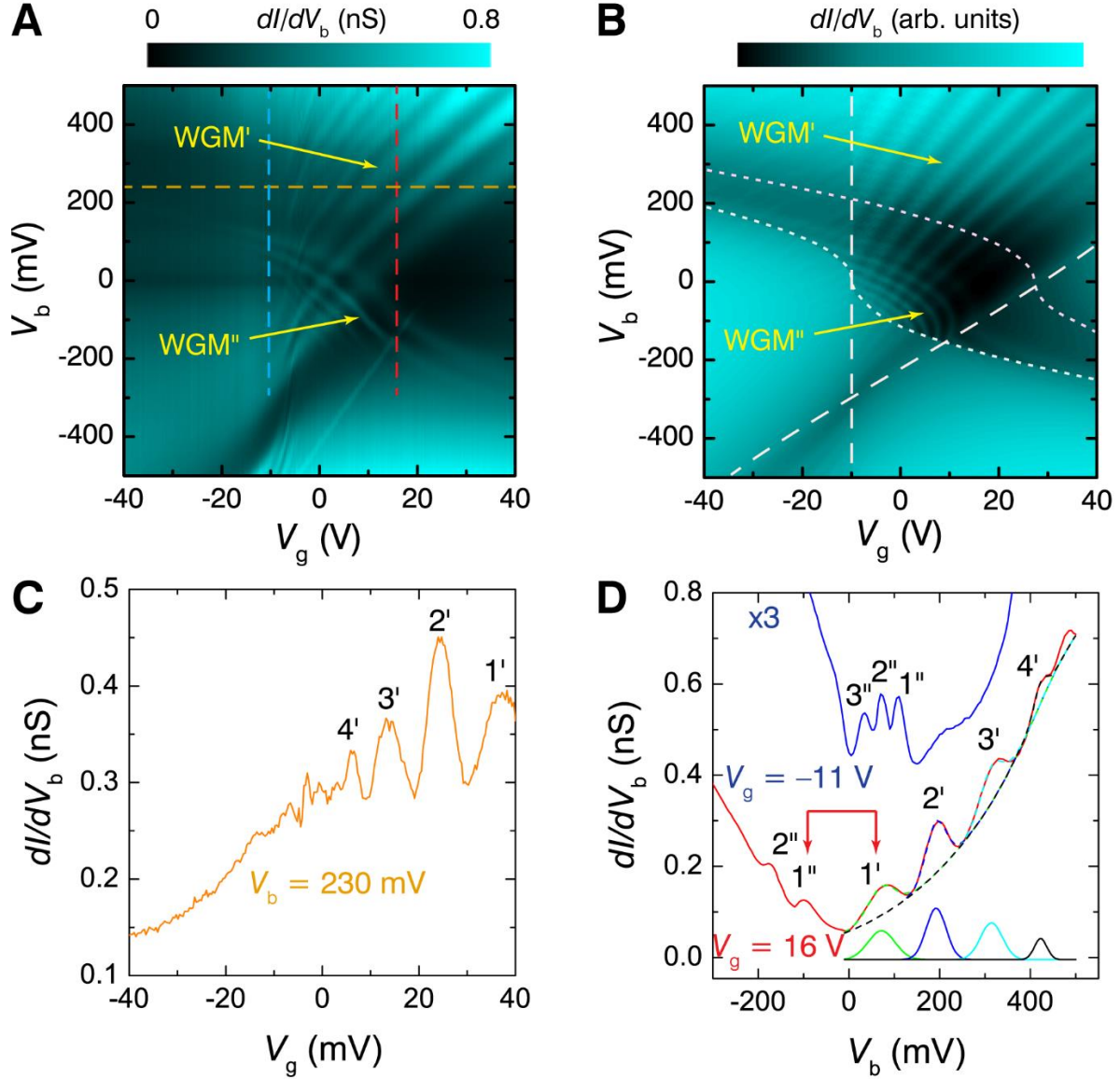

**Fig. 2. Confined electronic states probed by STM measurements.** (**A**) Differential tunneling conductance, $dI/dV_b$, map for a single-layer graphene device as a function of sample bias, $V_b$ and back gate voltage, $V_g$. The two fans of interference features, marked WGM' and WGM", originate from WGM resonances in the density of states, see text. (**B**) Interference features in $dI/dV_b$ calculated from the relativistic Dirac model. The features WGM' and WGM" in the $(V_g,V_b)$ map originate, respectively, from the conditions $\varepsilon = \mu_0$ and $\varepsilon = \mu_0 + eV_b$, see text. The boundaries of WGM' (WGM") regions are marked by dashed (dotted) white lines. (**C**) $dI/dV_b$ spectra taken along the horizontal line in (A) at $V_b = 230$ mV. (**D**) $dI/dV_b$ spectra taken along the two vertical lines in the map in (A) at $V_g = 16$V (red line) and at $V_g = -11$V (blue line, scaled x3 and offset for clarity), (see text for discussion). The four peaks at positive bias at $V_g = 16$V are fit to Gaussian functions, with the fits shown in the lower right of the figure. The peaks labeled 1",2",3"… correspond to WGM resonances probed at the energy $\varepsilon = \mu_0 + eV_b$, while those labeled 1',2',3'…, are the same WGM resonances probed at the Fermi-level $\varepsilon = \mu_0$, giving rise to the WGM" and WGM' fringes in the gate maps, respectively. The resonance spacing of order 40 mV, translates into a cavity radius of 50 nm, using the relation $\Delta\varepsilon = \pi\hbar v_F/r$ (see text).

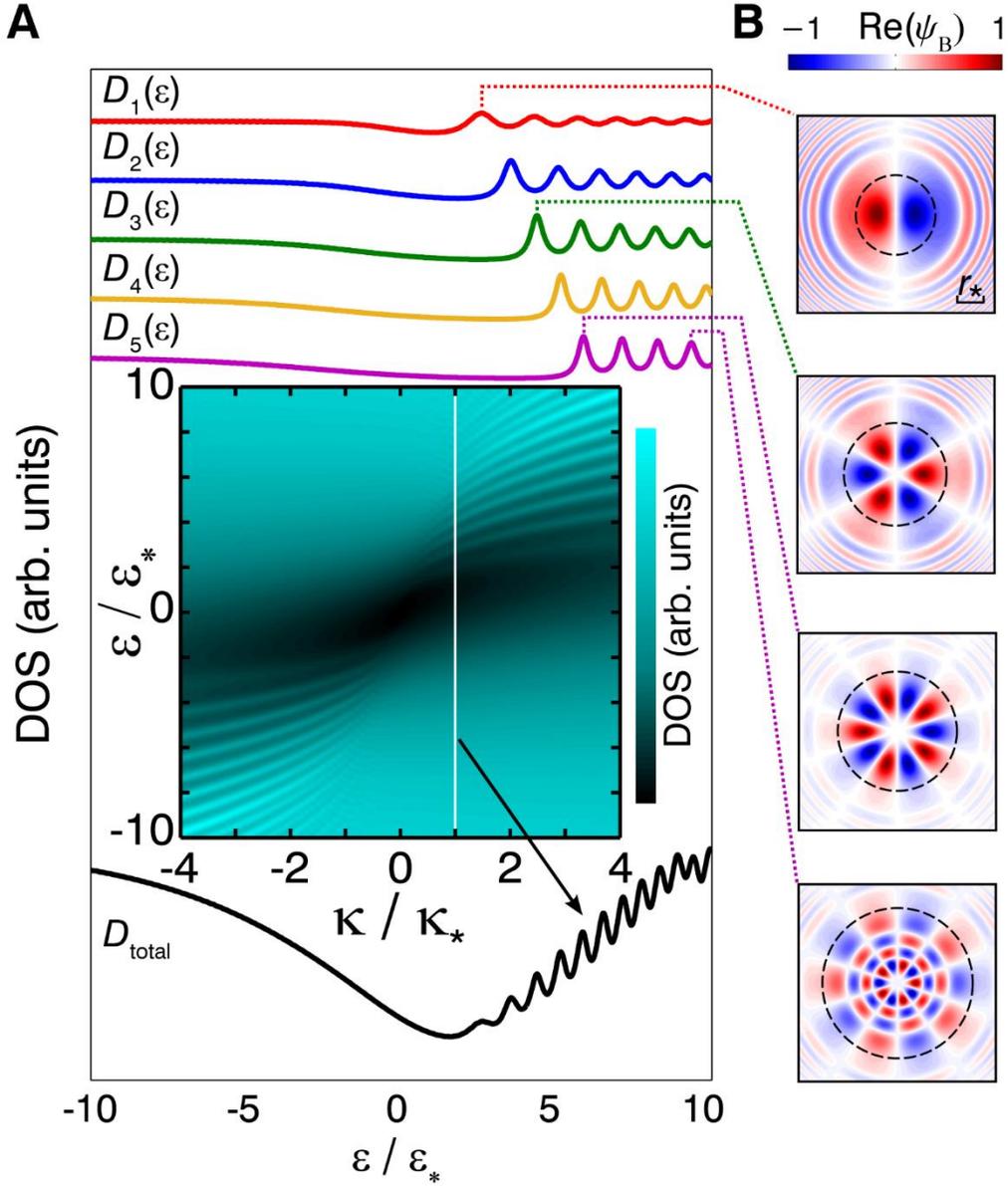

**Fig. 3. Contributions of the WGM resonances with different *m* to the DOS for the relativistic Dirac model.** (**A**) Colored curves represent partial-*m* contributions from angular momentum values *m*=1,2,3,4,5, for a confining potential $U(r) = \kappa r^2$ with curvature value $\kappa = \kappa_* = \varepsilon_*/Rd$. Different curves are offset vertically for clarity. The inset shows the total DOS *vs*. particle energy $\varepsilon$ and the curvature $\kappa$ (see text). Black curve shows the total DOS trace along the white line. (**B**) The Dirac wavefunction for different WGM states, Eq. (1). Spatial structure is shown for several resonances in the partial DOS (black dashed circles mark the *pn* junction rings). The quantity plotted, $\text{Re}(\psi_B)$, is the same as in Fig.1B. The lengthscale $r_* = \sqrt{Rd}$ (the same in all panels) is marked. Note the confinement strength increasing with *m*.

# Supporting Online Material for

## Creating and Probing Electron Whispering Gallery Modes in Graphene

Yue Zhao[1,2*], Jonathan Wyrick[1*], Fabian D. Natterer[1*], Joaquin F. Rodriguez-Nieva[3*], Cyprian Lewandowski[4], Kenji Watanabe[5], Takashi Taniguchi[5], Leonid S. Levitov[3], Nikolai B. Zhitenev[1†], and Joseph A. Stroscio[1†]

[1]Center for Nanoscale Science and Technology, National Institute of Standards and Technology, Gaithersburg, MD 20899, USA
[2]Maryland NanoCenter, University of Maryland, College Park, MD 20742, USA
[3]Department of Physics, Massachusetts Institute of Technology, Cambridge, MA 02139, USA
[4]Department of Physics, Imperial College London, London SW7 2AZ, UK
[5]National Institute for Materials Science, Tsukuba, Ibaraki 305-0044, JAPAN
*These authors contributed equally to this work
†Corresponding authors. E-mail: nikolai.zhitenev@nist.gov, joseph.stroscio@nist.gov

**This file includes:**

# I. Device Fabrication and Experimental Setup

The graphene on *h*-BN devices are fabricated for STM measurements using a dry transfer method described in Ref. (*29*). Particular attention is given to fabrication details to ensure clean graphene surfaces for tunneling spectroscopic measurements. First, *h*-BN flakes are exfoliated onto highly *p*-doped Silicon wafers (0.001-0.005 Ωcm in resistivity) with 285 nm of dry thermal oxide using adhesive tape. We select *h*-BN flakes with 15 to 30 nm thickness to optimize the screening capability of *h*-BN crystals from substrate impurities and visibility of the graphene flakes that will be stacked later in an optical microscope. Atomic force microscopy imaging is used to gauge their thickness and surface flatness. We then directly exfoliate graphene pieces onto a transparent supporting stack consisting of glass slide/Polydimethylsiloxane (PDMS)/Tape/Methyl methacrylate (MMA EL 11)/polymethyl methacrylate(PMMA A8). Raman spectroscopy measurements are used to identify the graphene layer numbers. After transferring graphene pieces onto *h*-BN flakes with a carrier film of MMA/PMMA, Cr(5nm)/Pd(15nm)/Au(30nm) contacts are deposited by multi-step standard electron-beam lithography, including a fan out of radial guidelines for STM navigation, as shown in Fig. S1A. Annealing in Ar/$H_2$ flow at 340 ºC of the *h*-BN flake and the final device ensures a clean graphene surface free of contamination. The device used in this manuscript consist of single layer (SLG), bilayer (BLG), and multilayer (MLG) graphene regions, as shown in Fig. S1B.

After the device is inserted into the ultra-high vacuum system, it is heated to 100 ºC and then inserted into the STM. Tunneling measurements were carried out with a custom built low temperature instrument, operating at 4.3 K and a base pressure lower than $5 \times 10^{-9}$ Pa (*30*). Prior to cooling to 4 K, the probe tip is aligned onto the graphene device area at room temperature (Fig.

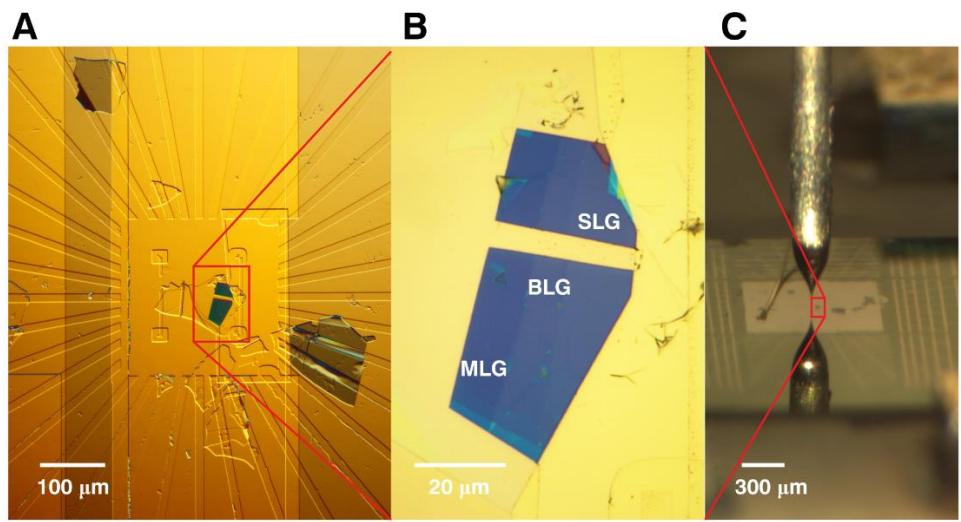

**Fig. S1. Graphene device fabrication.** Optical images of the graphene device at (**A**) 20X, (**B**) 150X magnification, and (**C**) inside the STM module showing the probe tip aligned with the active area of the device.

S1C) and maintains this alignment to within ≈10 μm after cooling. STM topographic images show a clean and defect free graphene surface with the honeycomb lattice structure, as shown in Fig. S2. Tunneling bias and back gate voltage dependent differential conductance (*dI/dV*) measurements were recorded by adding a sinusoidal voltage to the sample bias at 203.5 Hz. Measurements were made before and after the tip work-function was modified by backfilling the vacuum chamber with molecular deuterium (≈$5 \times 10^{-5}$ Pa for about 70 min) with the tip held atop the Au electrode. The dosing with $D_2$ had no other impact in measurements, neither in topography, nor did we observe any signatures of rotational excitations in inelastic tunneling spectroscopy (*31-33*).

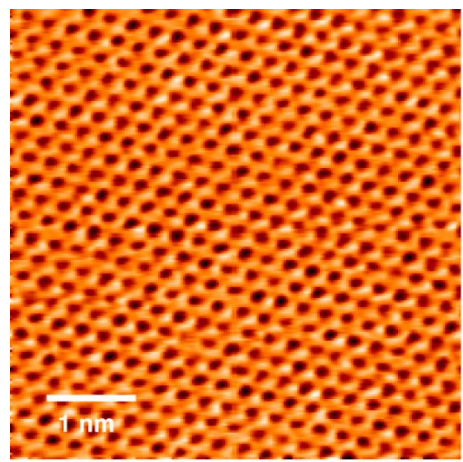

**Fig. S2. STM topographic image of the graphene lattice.** Tunneling parameters I= 100 pA, $V_b$ = 300 mV.

## II. Back Gate Mapping Tunneling Spectroscopy of Graphene Devices

Figure S3 shows the band structure alignment for different tunneling conditions. The contact potential, or work function difference between graphene and the metal probe tip leads to an electric field acting on graphene and results in a doping of the graphene layer when they are brought into tunneling distances, as shown in Fig. S3B. Tunneling occurs for the overlapping density of states between the Fermi levels of tip and sample, separated by the applied tunneling bias $V_b$, (Fig. S3, C and D). The WGM resonances are observed twice in the differential tunneling spectra, due to the electric field gating of the graphene by the sample-tip potential difference, as schematically shown in Figs. S3, C and D. First, the standard tunneling channel is observed at the lower tunneling barrier energies, corresponding to tunneling out of the graphene at negative sample bias (Fig. S3C), or out of the tip states at positive sample bias (Fig. S3D). This tunneling channel gives rise to the dispersive WGM" resonances displayed in Fig. 2A, which follow an inverted "S" shape like normal graphene states. The WGM' fan appears, when resonances become filled (Fig. S3C) or emptied (Fig. S3D) by field effect gating of graphene by the probe tip. Here the resonances coincide with the lower Fermi level of the tip or graphene, and give a peak in the differential conductance as the resonance is filled or emptied. This second channel gives rise to the "S" shaped WGM' fan in Fig. 2A. The larger energy spacing of the levels of the WGM' compared to the WGM" levels is due to the gating efficiency, or lever arm, $\eta$, in Eq. (6).

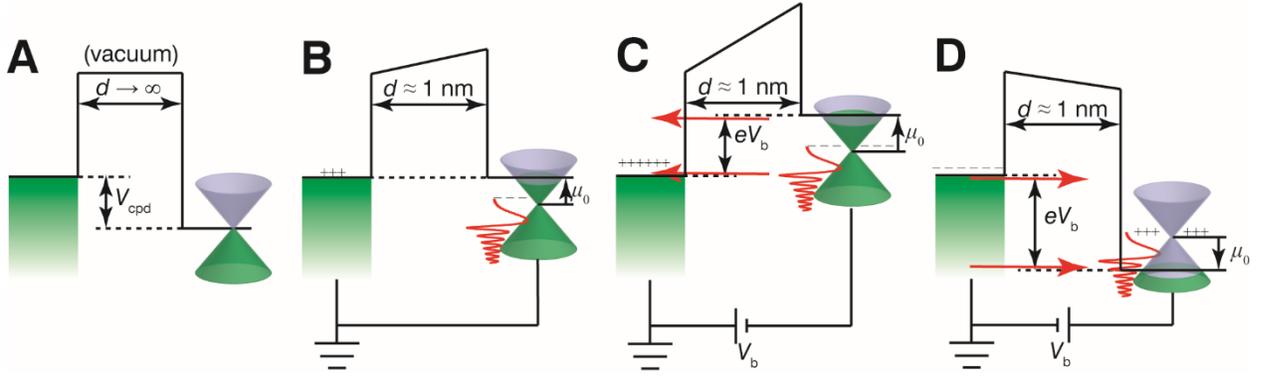

**Fig. S3. Band structure schematics showing band alignment between the STM probe tip and graphene sample**. (**A**) Large separation between probe and sample, (**B**) grounded and in tunneling proximity, (**C**) negative, and (**D**) positive sample biased tunneling conditions. Here, $V_{cpd}$ is the contact potential difference between graphene and the tip. The +,− signs indicate the charge polarities in the tip and in graphene. The two red arrows in (C) and (D) show that resonances in the density of states can appear twice in the differential tunneling conductance measurements and appear as the interference fans, WGM' and WGM". The WGM' fans appear when a resonance becomes filled and aligned with the tip Fermi level at negative sample bias (C), or when the resonance is empty and aligned with the graphene Fermi level at positive sample bias (D). These two cases produce the WGM' interference fans in the lower left and upper right of the ($V_b$, $V_g$) gate maps, as shown in Figs. 2, and Fig. S5. In contrast, the WGM" resonances appear from direct tunneling from the top most filled state in (C) or into the topmost empty state in (D).

The two tuning knobs, consisting of the back gate and tunneling sample bias voltages, allow the tunneling differential conductance to be mapped in the $(V_b, V_g)$ plane. As schematically shown in the main text Fig. 1A, the STM probe tip acts as a top gate thereby creating circular *pn* junctions in certain regions of the $(V_b, V_g)$ parameter space where they also reveal whispering gallery mode states. In Figure S4, we sketch the local and background doping as a function of position in the gate maps to delineate the regions of *pn* and *np* junctions. As we go from left to right at the top of the gate map we vary the doping from *p+p* to *pn* to *n-n*. Conversely, at the bottom of the gate map we vary the doping from *p-p* to *np* to *n+n*, as schematically shown by the circles in Fig. S4. A more detailed description of the potential profile as a function of $(V_b, V_g)$ is given in section III below.

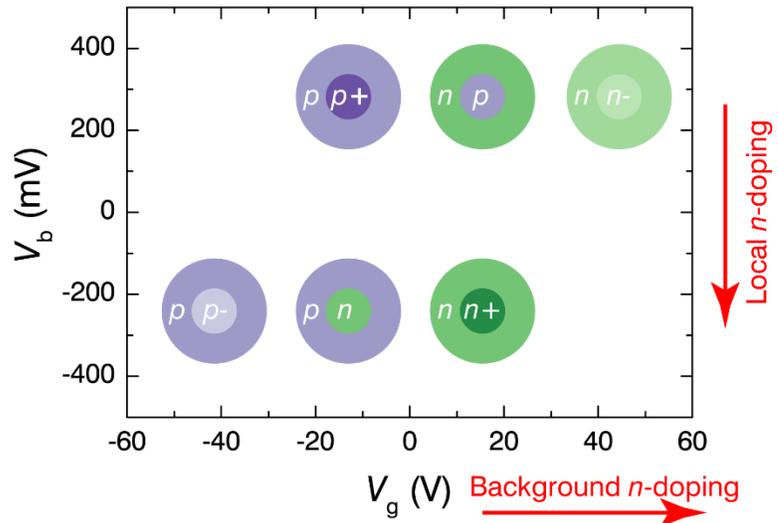

**Fig. S4. Schematic of the background vs. local field effect doping in the tunneling gate maps**. The background doping far from the probe tip is adjusted by the back gate electrode on the horizontal axis of the gate map. The local doping under the tip is adjusted by the tunneling bias on the vertical axis of the gate map. Both electrodes induce n-type doping by going towards the lower right corner of the gate map. The doping under the probe can be reversed relative to the background doping to create circular pn junction rings in the upper right and lower left of the gate map, as schematically indicated by the small disks representing the local doping, and the larger disks, the far field background doping.

The WGM' fans consist of an "S" shape in the $(V_b, V_g)$ maps (Fig. S5, A and B). The center of the "S" defines the charge neutrality point, which can be shifted along the $V_g$ axis due to impurity doping of the substrate, and along the $V_b$ axis due to work function differences between the probe tip and graphene. As seen in comparing the gate maps in Fig. S5, A and B, we shifted the center point of the WGM' fan to lower $V_b$ values in Fig. S5B by increasing the probe tip work function with $D_2$ adsorption. The fan center shifts downward ≈430 meV, consistent with a corresponding increase in work function from the dissociation of $D_2$ on the tip apex (*18*), and allows a greater portion of the WGM' fan resonances to be observed in the upper right portion of the gate map.

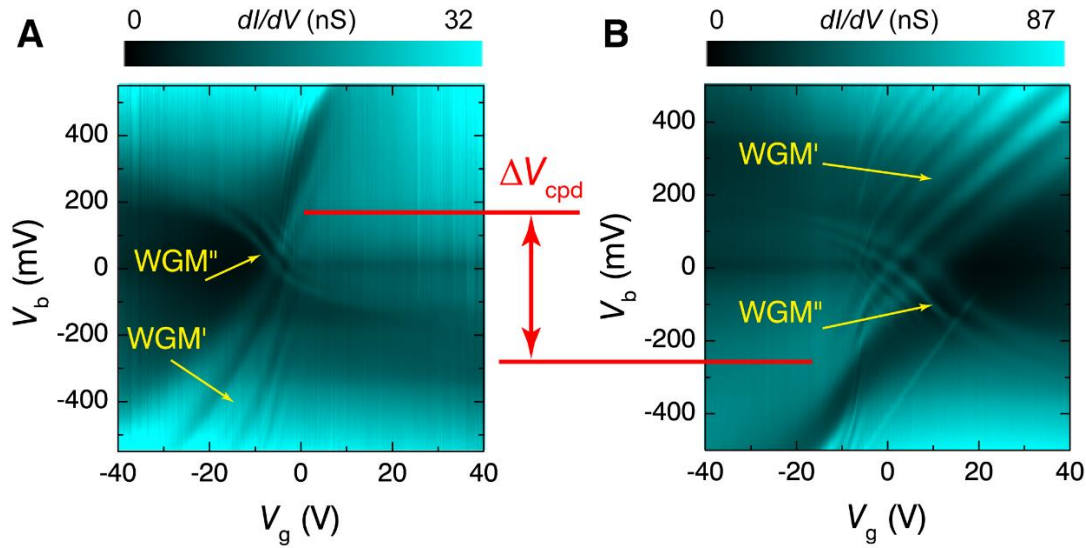

**Fig. S5. Shifting of the WGM fans by changes in the probe work function**. (**A**) Initial *dI/dV* gate map before changing the probe tip work-function. The WGM' fan is mainly seen in the lower left of the gate map, with a neutrality point at $\approx V_b = 160$ mV. To see a larger extent of the WGM' fan at positive tunneling bias requires a larger positive bias range (not shown). (**B**) *dI/dV* gate map after the probe tip was exposed to $D_2$. Note the WGM' fans shifted vertically down due to an increase in the tip work function from $D_2$ adsorption, with a neutrality point now at $\approx V_b = -270$ mV. The WGM' fans are now mainly seen in the upper right of the gate map. From the shift in the WGM' fan we estimate a work-function increase of the probe tip to be ≈430 meV.

## III. Tunneling Density of States and Electrostatics Modeling

To find the eigenstates of the Hamiltonian, Eq. (2), we use the finite difference method in the interval $0 < r < L$ with large enough $L$ and with a large repulsive potential at $r = L$ in order to confine the eigentstates. We use a finite range $-M \leq m \leq M$ for the azimuthal quantum numbers with the value $M$ chosen large enough to represent accurately the states in the energy range of interest. The *local* density of states beneath the tip, $D(\varepsilon) = \sum_m D_m(\varepsilon)$, used for modeling the conductance maps, Eq. (4), is calculated from the eigenvalues and eigenstates of the Hamiltonian, Eq. (2), as

$$D_m(\varepsilon) = \sum_{\nu=1}^{2N} \frac{\Gamma}{\pi} \cdot \frac{\langle |u_\nu(r=0)|^2 \rangle_\lambda}{(\varepsilon - \varepsilon_\nu)^2 + \Gamma^2}. \quad (S1)$$

Here, the broadening parameter value $\Gamma$ is chosen to be a few times the level spacing, $\Gamma \sim \hbar v/L$, $\nu$ labels the radial eigenstates of Eq. (2), and the average $\langle ... \rangle_\lambda$ denotes

$$\langle |u_\nu(r=0)|^2 \rangle_\lambda = \int_0^L dr' |u_\nu(r')|^2 \exp(-\lambda r'^2/2). \quad (S2)$$

The Gaussian in Eq. (S2) accounts for the finite size of the tunneling area due to the finite curvature radius of the STM tip and/or a residual asymmetry of the STM tip, both of which allow electrons to tunnel some distance off the tip center. Spurious states arising from the finite potential jumps at the boundaries, localized within a few lattice sites of $r = 0$ and $r = L$ and with energies roughly independent of κ, are excluded from the sum in Eq. (S1).

In our calculations, we use a system of size $L/r_* = 20$, where $r_*$ is defined as $r_* = \sqrt{Rd}$ [see Eq. (2) in main text]. The interval $0 < r < L$ was discretized with $N = 1200$ points, yielding an $2400 \times 2400$ matrix for each value of $\kappa$ and $m$. The azimuthal quantum number maximum value $M = 8$ was used. In Eq. (S1), we used the value of $\Gamma/\varepsilon_* = 0.2$ for broadening and $\lambda = 1$ for the Gaussian width parameter. The latter corresponds to an effective tunneling region of size $\lambda^{-1/2} r_* \sim 22$ nm, where $r_*$ is obtained for typical STM tip parameter values (see details below). Such values for the tunneling region size, while may seem to be a bit on the high side, are in fact not unreasonable: For instance, the off-center tunneling rate decays exponentially, governed by the WKB exponent $\sim \exp[-2d_t(r)\sqrt{2m_0 W}/\hbar]$, where $d_t$ is the local tip-graphene distance, $m_0$ is the effective electron mass and $W$ is the work function. The spread for electrons away from the tip center can be estimated from $d_t(r) \approx d(1 + r^2/2Rd)$, yielding an effective value of $\lambda = 2d\sqrt{2m_0 W}/\hbar \sim 5$ and thus a tunneling region of size $\lambda^{-1/2} r_* \sim 10$ nm (here we used the electron mass $m_0 \sim 9.1 \cdot 10^{-31}$ kg, $W \sim 1$ eV, and $d \sim 0.5$ nm). We expect fabrication-induced tip asymmetries to be in the same ballpark.

We use the Thomas-Fermi (TF) model to relate the parameters $(\varepsilon, \kappa)$ in the Hamiltonian, Eq. (2), to the experimental knobs $(V_b, V_g)$. Since the graphene-tip distance $d$ is much smaller than the tip radius $R$, $d \ll R$, the tip-induced carrier density in graphene $\Delta n = \text{sgn}(\mu)\mu^2/\pi(\hbar v_F)^2 - n_\infty$ can be modeled as

$$\frac{\text{sgn}(\mu)\mu^2}{\pi(\hbar v_F)^2} - n_\infty = -\frac{e(V_b - V_{\text{cpd}}) + \mu}{4\pi e^2 (d + r^2/2R)}. \quad (S3)$$

Here, μ is the position-dependent Fermi energy taken relative to the Dirac point, $n_\infty$ is the carrier density far from the STM tip (controlled by the gate potential), and $V_{\text{cpd}}$ is the contact potential difference between the tip and neutral graphene. The right-hand side of Eq. (S3), is obtained from a parallel-plate capacitor model with slowly varying interplate-distance $d_t(r) \approx d + r^2/2R$, and neglecting the higher-order corrections due to the curvature of the field lines. This yields a screening length scale $\sqrt{2Rd} \ll R$, as described in the main text. The carrier density far from the tip is controlled by the gate potential as $n_\infty = \epsilon e(V_g - V_g^0)/4\pi e^2 d_g$, where $\epsilon$ and $d_g$ are the dielectric constant and thickness of the dielectric substrate, respectively, and $V_g^0$ accounts for residual doping in graphene at $V_g = 0$.

We analyze the nonlinear screening problem accounting for the fact that the spatially varying carrier density may change polarity near the STM tip. This is done with the help of the self-consistent TF model, Eq. (S3), treated as an algebraic equation for µ. Parameters of interest for the microscopic model, $\mu_0$ and κ, are obtained directly from Eq. (S3). The value of $\mu_0 = \mu(r = 0)$ is given by

$$\mu_0(V_b, V_g) = \frac{\alpha - \sqrt{\alpha^2 + 4|\beta(V_b, V_g)|}}{2\text{sgn}[\beta(V_b, V_g)]}, \quad (S4)$$

where β and α are

$$\beta(V_b, V_g) = \alpha\left[e(V_b - V_{\text{cpd}}) - \frac{\epsilon d}{d_g}e(V_g - V_g^0)\right], \quad \alpha = \frac{(\hbar v_F)^2}{4e^2 d}. \quad (S5)$$

The parameter κ in our parabolic approximation for $U(r)$ is calculated as $\kappa = -\mu''(0)/2$ and giving

$$\kappa(V_b, V_g) = -\frac{\alpha}{2Rd}\frac{e(V_b - V_{\text{cpd}}) + \mu_0(V_b, V_g)}{\sqrt{\alpha^2 + 4|\beta(V_b, V_g)|}}. \quad (S6)$$

Importantly, we note that the potential term in our model Hamiltonian, $U(r) = \kappa r^2$, is unbounded, whereas in reality it has maximum value $U(|x| \to \infty) = \mu_0 - \mu_\infty$ (see Figs. S6 and S7). This constraints the regions in the $(V_g, V_b)$ map where WGMs are expected to occur, as discussed in the main text and below. In particular, states with energy $|\varepsilon| > |\mu_0 - \mu_\infty|$ do not have a bipolar character (*i.e.*, do not form a *pn* junction). In order to include the finite value for $U(|r| \to \infty)$ in plotting Fig. 2B, we treat the value of κ as given by Eq. (S6) when $|\varepsilon| \leq |\mu_0 - \mu_\infty|$, with $\text{sgn}(\varepsilon) = \text{sgn}(\mu_0 - \mu_\infty)$, and as $\kappa = 0$ otherwise.

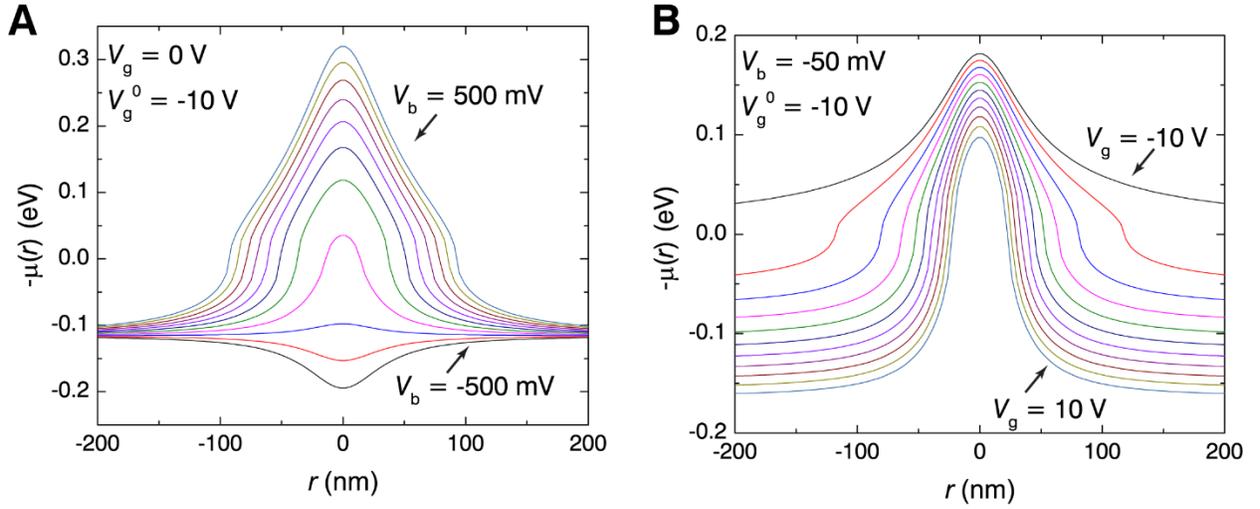

**Fig. S6. Radial Fermi energy profile obtained from the TF model (see text).** The profile was calculated from Eq. (S3) (**A**) as a function of bias voltage $V_b$ varying from $-500$ mV to 500 mV in steps of 100 mV for a fixed gate voltage $V_g = 0$, and in (**B**) it was calculated as a function of gate voltage $V_g$ varying from -10 V to 10 V in steps of 2 V for fixed bias voltage $V_b = -50$ mV. In both panels, we included the effect of residual doping in graphene by adding a gate potential offset of $V_g^0 = -10$ V, see text. The tip-induced circular *pn* and *np* junctions appear when the potential profile crosses 0 eV. The *pn* junction radius varies between 0 and 100 nm.

Figure S6A illustrates the influence of the sample-tip bias acting as a local top gate; the potential profile sign reverses from local hole doping at large positive bias to electron doping at strong negative bias. Figure S6B, likewise, illustrates the impact of the back gate potential that is mainly to modify the tails of the local potential induced by the tunneling tip. For parameter values of our TF model, we use $R \approx 1$ μm, $d \approx 0.5$ nm, and $V_{cpd} \approx -0.3$ V. This yields characteristic length and energy scales $r_* \approx 22$ nm and $\varepsilon_* \approx 30$ meV, respectively. The backgate in our device is separated from graphene by a dielectric of thickness $d_g \approx 300$ nm and dielectric constant $\epsilon \approx 5$. We therefore introduce a gate potential offset of $V_g^0 = -10$ V to match the residual doping observed in our experiments. Using these parameters and Eq. (S3), Fig. S6A displays the spatially varying Fermi energy $\mu(r)$ for different values of $V_b$ and a fixed gate potential $V_g = 0$. Figure S6B shows the Fermi energy $\mu(r)$ profile vs. gate potential $V_g$ for fixed $V_b = -50$ mV.

In Fig. S7 we plot the shape of the spatially varying electrostatic potential in the $(V_g, V_b)$ gate map space (blue lines). The relative alignment of the Fermi level under the tip $\varepsilon = \mu_0$ (orange lines) and the sample bias energy $\varepsilon = \mu_0 + eV_b$ (red lines) with respect to the potential profile is indicated for every $(V_g, V_b)$ scenario. One can qualitatively understand where the WGM' and WGM'' resonances will appear by examining when the potential intersects the orange and red lines (i.e., the appearance of a *pn* junction for the electronic state in question): the WGM' resonances in Fig. 2 of the main text can appear when the potential profile crosses the $\varepsilon=\mu_0$ lines (orange), while the WGM'' resonances emerge when the potential profile intersects the sample bias $\varepsilon = \mu_0 + V_b$ (red lines). Both scenarios occur inside the regions of the $(V_g, V_b)$ map indicated with dashed (WGM') and dotted (WGM'') lines in Fig. 2B of the main text, and in Fig. S7. Additionally, the dependence on gate voltages of the *pn* ring size is also apparent in Figs. S6 and S7. For example, for the different values of bias and gate voltages considered in Fig. S6, a state at the Fermi Level $\varepsilon = \mu_0$ will have classical turning points where the potential profile crosses zero.

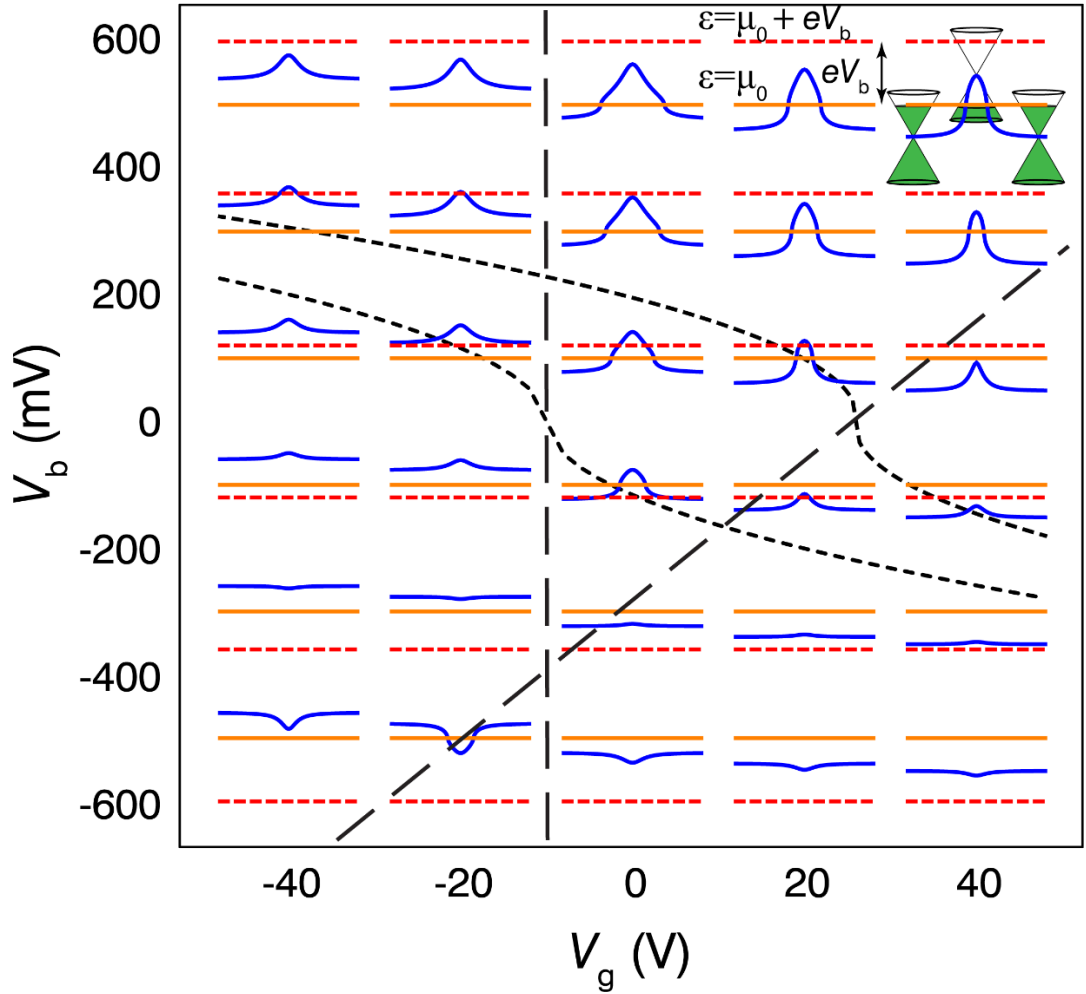

**Fig. S7. Variation of the electrostatic potential profile in the $(V_g, V_b)$ plane.** The potential profile is shown by the blue curves calculated as a function of $(V_g, V_b)$. The level being probed by STM is indicated with an orange line for $\varepsilon = \mu_0$ and with a red line for $\varepsilon = \mu_0 + eV_b$. Circular *pn* an *np* junctions induced by the tip gating can be seen when the potential profile crosses the $\varepsilon = \mu_0$ lines. The solid dotted lines denote the boundaries of the WGM'' resonances, which can be seen for locations when the blue potential curves cross the red $eV_b$ lines (these boundary lines are also indicated in Fig. 2B of the main text). Similarly, dashed black lines indicate the regions where the WGM' resonances are seen in the experimental gate maps, which occurs when the potential curves cross the (orange) energies $\varepsilon = \mu_0$. The lateral spatial extent for each blue curve ranges from −200 nm to 200 nm.